\documentclass[10pt,aps,twocolumn,showpacs,superscriptaddress,floatfix,nofootinbib]{revtex4-2}

\pdfoutput=1
\usepackage[utf8]{inputenc}
\usepackage[english]{babel}
\usepackage[T1]{fontenc}
\usepackage{csquotes}
\usepackage{amsmath,amsfonts}
\usepackage{enumitem}

\usepackage[dvipsnames]{xcolor}
\usepackage[colorlinks=true]{hyperref}
\usepackage[capitalise]{cleveref}

\usepackage{listings}

\usepackage{multirow}
\usepackage{graphicx}
\usepackage{subcaption}
\usepackage{tikz}

\graphicspath{{figures/}}

\usepackage{bm} 

\usepackage[normalem]{ulem} 

\definecolor{ao(english)}{rgb}{0.0, 0.5, 0.0}
 
\DeclareRobustCommand{\todelete}[1]{}

\begin{document}
\title{Excitons in solid molecular hydrogen from quantum Monte Carlo}

\author{Edgar Josué Landinez Borda}
\altaffiliation{Present address: Forschungszentrum Jülich GmbH, Wilhelm-Johnen-Straße, 52428 Jülich, Germany}
\affiliation{MESA+ Institute for Nanotechnology, University of Twente, P.O. Box 217, 7500 AE Enschede, The Netherlands}
\author{Ravindra Shinde}
\affiliation{MESA+ Institute for Nanotechnology, University of Twente, P.O. Box 217, 7500 AE Enschede, The Netherlands}
\author{Claudia Filippi}
\affiliation{MESA+ Institute for Nanotechnology, University of Twente, P.O. Box 217, 7500 AE Enschede, The Netherlands}
\author{Markus Holzmann}
\affiliation{Univ. Grenoble Alpes, CNRS; LPMMC, 38000 Grenoble, France}
\author{Saverio Moroni}
\affiliation{CNR-IOM DEMOCRITOS, Istituto Officina dei Materiali and SISSA Scuola Internazionale Superiore di Studi Avanzati, Via Bonomea 265, I-34136 Trieste, Italy}
\begin{abstract}
We use quantum Monte Carlo (QMC) methods to investigate the lowest neutral 
excitation of solid molecular hydrogen in phase-I under increasing compression, 
with emphasis on structural and response properties.
We employ QMC wave functions for the excited state that feature
only a single orbital substitution, or a multideterminant
expansion including every single particle-hole transition,
and compare the results with the Bethe-Salpeter
equation (BSE) and time-dependent density-functional theory (TDDFT). 
We examine a number of nuclear configurations representing
thermal and quantum fluctuations.
The effect of the multideterminant expansion is 
marginal on the QMC excitation energies but crucial on the
exciton localization and related quantities, irrespective of 
full optimization and imaginary time projection.
In the strong insulating regime at low density, 
multideterminant QMC results match and in some cases surpass 
the degree of localization afforded by BSE. 
A similar trend is observed for transition dipole moments and 
excitation-induced forces, with generally good agreement between 
QMC and BSE but marked discrepancies for nuclear configurations
where localization differs significantly.
TDDFT generally gives a more extended exciton with correspondingly
weaker transition dipole moments.
These differences decrease upon compression, as the system approaches 
a semiconducting regime where multiple particle-hole transitions are 
less important.

\end{abstract}
\maketitle

\section{Introduction}

Despite its chemical simplicity, solid hydrogen displays complex behavior
under pressure and remains challenging for first-principles
electronic-structure methods~\cite{Mcmahon_2012,Bonitz_2024}. 
An accurate description of the different phases in general requires both
electronic correlation and nuclear quantum and thermal effects, which 
produce substantial fluctuations of the nuclear configurations.

A recent theoretical study~\cite{Gorelov_2024} of the electronic excitation spectrum
of phase-I hydrogen over pressures from 5 to 90 GPa showed that excitonic effects
are essential for describing the measured inelastic X-ray scattering
spectra~\cite{Li_2021}. Across this range, phase-I hydrogen evolves
from a wide-gap molecular insulator toward semiconductor-like behavior with
increasing compression, accompanied by a reduction of the exciton binding energy
from about 2 to 0.5 eV~\cite{Gorelov_2024}. These results were obtained
from the Bethe-Salpeter equation  
(BSE) and supported by quantum Monte Carlo (QMC) calculations.
However, in QMC the excitonic effects were addressed only indirectly,
relying on an assumption about the extended or localized nature of the exciton based on its BSE wave function.

In the present work, we extend the QMC~\cite{Foulkes2001,Luchow2011,Austin2012,KolorencMitas2011} description of excitations,
often limited to the calculation of energies, to a wider spectrum of physical properties
related to a key feature of the exciton, its localization. 
To this end, we examine the change in the electronic density
and static structure factor, the transition dipole moment, and 
the change in the forces acting on the nuclei upon excitation. 

Extensive comparisons of our QMC results with those obtained from the BSE~\cite{Strinati1988,RohlfingLouie2000,OnidaReiningRubio2002} 
and time-dependent density-functional theory (TDDFT)~\cite{RungeGross1984,Petersilka1996,Casida1995,TDDFT2}, two widely used
approaches to electronic spectra, support the reliability of our self-contained description of the exciton while revealing significant differences between
the methods in some cases.

The pivotal element to correctly
describe localization turns out to be the use of multideterminant expansions in the excited state wave function.
The multideterminant nature of excitation in solids has been addressed in 
several QMC calculations, for instance using configuration state functions (CSFs)
to represent the symmetry of multiplet states~\cite{needs,makkonen}, 
a small set of active orbitals to enforce a pre-established spin structure~\cite{wagner-spin}
or to control occupations and spin couplings~\cite{KleinerEtAl2025,KleinerWagner2026},
and selected- or full-configuration interaction (CIPSI or FCI)~\cite{benali21,benali24,alavi}
that can provide extremely accurate benchmarks for sufficiently small system size and basis set.

Here we consider instead a configuration-interaction singles (CIS) expansion.
This approach is system-agnostic, unlike selection based on symmetries or intuition
on the relevant orbitals to include, and scalable to large system size, unlike CIPSI or FCI.
While CIS is not expected to give relevant improvement to the variational energy of molecules
and finite clusters~\cite{cis},
it is appealing on physical grounds for the description of excitations in the solid state 
because it parallels the representation of the 
exciton wave function as a superposition of multiple particle-hole transitions at the core
of the very successful BSE and TDDFT.

We find indeed that QMC excitation energies change little upon introducing the CIS expansion,
while the spatial properties can change dramatically. 
The comparison with BSE and TDDFT is carried out for several nuclear
configurations affected by thermal and quantum fluctuations, for which the
degree of electronic localization in the excited state can vary significantly
depending on the 
theory level.
At low density, some nuclear configurations show a clear progression toward stronger
localization from the the description with a single-CSF excited wave function to results from TDDFT, BSE and CIS, with BSE generally closer to the 
multideterminant QMC calculations. For other configurations, strong localization appears only
with the CIS wave function. These differences become smaller
upon compression. 


\section{Methods}
\label{sec:methods}

We study excitons in solid molecular hydrogen in phase I at room temperature  
for three compressed densities, $\rho/\rho_0 = 3.15$, 4.47, and 8.48, 
where $\rho_0 = 0.0396$~g/cm$^3$, 
corresponding to pressures of approximately 5, 10, and 90~GPa, respectively~\cite{Gorelov_2024}. 

Within the Born-Oppenheimer approximation, we represent
thermal and quantum fluctuations of the hydrogen molecules considering
for each density a number of nuclear configurations 
(respectively 10, 20 and 19) sampled with $NVT$-Smart Monte Carlo
in Ref.~\cite{Gorelov_2024}.
All electronic structure calculations in the present paper are performed 
on these static nuclear configurations with $\Gamma$-point sampling
for periodic simulation cells containing $N=96$ hydrogen atoms.

\subsection{QMC ground- and excited-state wave functions}
\label{sec:qmc_wf}

For a fixed nuclear
configuration $\mathbf{S}$, we determine the electronic ground state and the lowest singlet neutral excited
state, with energies $E_0({\bf S})$ and $E_1({\bf S})$ and wave functions $\Psi_0(\mathbf{R};\mathbf{S})$ and $\Psi_1(\mathbf{R};\mathbf{S})$,
respectively, where ${\bf R}=\{\mathbf{r}_1,\ldots,\mathbf{r}_N\}$ denotes the coordinates of the $N$ electrons. 
In the following, the dependence of the electronic wave functions on the nuclear coordinates
is left implicit.

We express $\Psi_0$ and $\Psi_1$ as Jastrow-Slater wave 
functions of the general form
\begin{equation}
\Psi_I({\bf R}) = J_I({\bf R}) \sum_{k=1}^{N_{\rm CSF}} c_{Ik} C_k[{\phi}_I]({\bf R}),
\label{eq:qmc_wf}
\end{equation}
with $I=0,1$.
The antisymmetric component of the wave function is determined by the
expansion in spin-adapted CSFs $C_k$ 
with coefficients $c_{Ik}$, where
${\phi}_I$ denotes the set of single-particle orbitals. 
Because the disordered configurations have no
spatial symmetry, each singlet spin-adapted CSF is a symmetric combination of 
two determinants obtained by the same spatial-orbital substitution 
in the two spin channels. 
All particle-hole pairs can in principle contribute to the expansion
as no substitution is excluded by symmetry. 
The Jastrow factor $J_I$ explicitly accounts for
electron-electron and electron-nucleus two-body short-range correlations. 
Each pair of particles contributes a factor $\exp(-u(r))$ to $J_I$, where
$r$ is the interparticle distance, and $u(r)$ is parametrized as 
\begin{eqnarray}
u(r)=\sum_{n=0}^{5} a_n r^n\left(1-\frac{r}{r_{\rm cut}}\right)^3\theta\left(1-\frac{r}{r_{\rm cut}}\right),
\label{ShortRangeJastrow}
\end{eqnarray}
with independent variational parameters $a_n$ for each state and for like- or unlike-spin electrons, and appropriate electron-electron or electron-nucleus
cusp conditions (here we drop indices of state, type of particles and spin).  
The cutoff distance $r_{\rm cut}$ is fixed at half the shortest
side of the simulation box.

To disentangle the effect on excitonic properties of individual 
factors within the general expression (\ref{eq:qmc_wf}),
we consider a hierarchy of six pairs $\{\Psi_0,\Psi_1\}_{X}$,
where {\em X} labels different forms and levels of optimization.

\begin{enumerate}[label=\arabic*)]
\item
The most basic level, {\em X=1CSF0}, corresponds to the 
Hartree-Fock (HF) Slater determinant for the ground state and a single CSF 
with HOMO-LUMO orbital substitution for the excited state, with
$J_I$ set to 1.

\item
In the pair labeled {\em X=J1CSF} we 
optimize both the single-particle orbitals and
the Jastrow factor for each state, maintaining
the single-CSF determinantal
structure of the previous item. 

\item
For {\em X=CIS0}, the ground state is the HF Slater determinant,
while the excited state is described by a CIS expansion
\begin{equation}
\sum_{k=1}^{N_{\rm CSF}} c_{1k} C_k[\phi_1]\to
\sum_{v=1}^{N/2} c_{1vc}\,C_{vc}[\widetilde{\phi}_1],
\label{eq:cis_wf}
\end{equation}
where $v$ and $c$ subscripts label occupied (valence) and
virtual (conduction) orbitals, respectively.
Here we use a singular-value decomposition of the CIS amplitude 
matrix to obtain a compact expansion~\cite{Martin2003,Mayer2007} of
just $N/2$ CSFs expressed in natural transition orbitals $\widetilde{\phi}_1$,
fully equivalent to the inclusion of all possible 
particle-hole substitutions in canonical orbitals.
Both Jastrow factors $J_I$ are set to 1.

\item
{\em JCIS0} denotes the inclusion of optimized Jastrow factors $J_I$.
Neither the orbitals nor the expansion coefficients are modified
with respect to the previous item.

\item
In {\em J}$_{LR}${\em CIS0}, the electron-electron Jastrow factors are 
supplemented by a long-range RPA contribution~\cite{Ceperley1978},
with reoptimization of the parameters $a_n$ pertaining to the
short-range part [Eq.~(\ref{ShortRangeJastrow})].
The RPA pair correlation reproduces the correct behavior in the 
long-wavelength limit, which is difficult to achieve by variational 
optimization because it has only a minor effect on the energy.

\item
Finally, in the {\em JCIS} wave functions we optimize
all the (short-range) Jastrow parameters, one-particle orbitals 
and CIS expansion coefficients.
For the ground state, the optimization of the determinantal
part of the wave function starts from the HF determinant, but
it acquires multideterminant character as the optimization is
carried out concurrently with the CIS excited state. However,
as expected based on the small energetic contribution
of the CIS, the final ground-state expansion has exceedingly small 
coefficients, and the description of the ground-state remains
effectively monodeterminantal.

\end{enumerate}

We perform VMC simulations for all the above wave functions,
as well as fixed-node DMC simulations~\cite{umrigar93} for {\em J1CSF} and 
{\em JCIS}, which have different nodal surfaces.

\subsection{Exciton characterization}
\label{sec:exciton_characterization}

Beyond the excitation energy, we examine the character and spatial extent of the exciton.
In a periodic system, a localized electron-hole excitation can be represented as a coherent superposition 
of particle-hole transitions involving different Bloch states. In the finite simulation cell, 
the CIS expansion generates such a superposition over occupied-to-virtual orbital substitutions.
Disorder in the nuclear configurations can also induce localization but the comparison of 
the single-CSF and CIS-based QMC wave functions isolates the effect of 
the multideterminant expansion on the spatial structure of the exciton.

The particle-hole character of the CIS excitation is analyzed in the natural-orbital representation
described above. 
We use the largest squared coefficient in the CIS expansion, $c_{\rm max}^2$, to quantify the extent
to which the excitation is dominated by a single particle-hole pair.
Values approaching one correspond to an essentially single-pair excitation, while smaller 
values indicate contributions from several particle-hole pairs. For comparison, we also consider 
the largest squared transition amplitude of the BSE excitation as an analogous indicator.

We describe the redistribution of electronic charge upon excitation through the density difference between the
excited and ground states,
\begin{equation}
\Delta\rho({\bf r})=\rho_1({\bf r})-\rho_0({\bf r}),
\label{eq:delta_rho}
\end{equation}
which shows directly where electronic density is depleted or accumulated. 
The spatial 
extent of the excitation is also probed through the static structure factor,
through the difference
\begin{equation}
\Delta S({\bf k}) =S_1({\bf k})-S_0({\bf k}),
\label{eq:delta_sk}
\end{equation}
where 
\begin{equation}
S_I({\bf k})=\frac{1}{N}\bigl<(\rho_{\bf k}-\langle\rho_{\bf k}\rangle_I)(\rho^*_{\bf k}-\langle\rho^*_{\bf k}\rangle_I)\bigr>_I
\label{eq:sk_def}
\end{equation}
is the static structure factor of state $I$, and $\langle\cdot\rangle_I$
denotes the expectation value on $\Psi_I$. Note that we use the
deviation of the density fluctuation operator
$\rho_{\bf k}=\sum_j {\rm exp}(-i{\bf k}\cdot{\bf r}_j)$ from its average,
rather than $\rho_{\bf k}$ itself. The resulting $\Delta S({\bf k})$ is
smoother and it has the same long-wavelength behavior.
For a neutral excitation localized within the simulation cell, 
the long-wavelength density fluctuations of the ground 
and excited states become identical, leading to
\begin{equation}
\lim_{k\rightarrow0}N\Delta S(k)=0.
\label{eq:localization}
\end{equation}
A nonvanishing extrapolation of $N\Delta S(k)$ to $k=0$ instead indicates 
that the excitation remains extended on the scale of the 
simulation cell~\cite{Gorelov_2024}.  
Because the calculations are carried out in a finite 96-atom cell, 
the $k\rightarrow0$ limit cannot be reached explicitly. 
We therefore infer localization from the behavior of $\Delta S(k)$ at 
the smallest accessible wave vectors and do not attempt a 
thermodynamic-limit extrapolation.

We characterize the optical response through the electronic transition dipole moment (TDM), 
which enters the optical absorption spectrum.
For exact eigenstates of the Hamiltonian $\hat H$, 
the transition dipole matrix element in atomic units can be evaluated as~\cite{baroni}
\begin{equation}
{\boldsymbol\mu}_{01}=
\frac{\langle\Psi_0|[H,\sum_j{\bf r}_j]|\Psi_1\rangle}{E_0-E_1}=-\frac{\langle\Psi_0|\sum_j\nabla_j|\Psi_1\rangle}{E_0-E_1}
\label{eq:tdm}
\end{equation}
%
%
%
The second equality, which holds for local potentials as used here,
avoids the direct evaluation of the position operators ${\bf r}_j$, ill-defined in periodic boundary conditions.
We calculate ${\boldsymbol\mu}_{01}$ in VMC. Since $\Psi_0$ and $\Psi_1$ are not exact eigenstates of $H$,
the first equality in Eq.~(\ref{eq:tdm}) holds only approximately. However, values obtained with
trial functions of the same nature but very different quality such as {\em CIS0} and {\em JCIS} agree to better than 
$\approx$3\%, much less than the discrepancies observed between VMC and BSE or TDDFT results (see below).
Therefore the comparison of different methods is not undermined by the bias incurred using Eq.~(\ref{eq:tdm})
within VMC.
We employ the symmetrized estimator $(\mu_{01}+\mu_{10})/2$
using that $\mu_{01}=\mu_{10}^*$ and that the wave functions are real.
The estimates of $\mu_{01}$ and $\mu_{10}$ have a strong statistical correlation, resulting in a
reduction of the variance by a factor of $\approx$50.

Finally, the response of the nuclei to the electronic excitation is characterized through the force difference
\begin{equation}
\Delta{\bf F}_A={\bf F}^{(1)}_A-{\bf F}^{(0)}_A ,
\label{eq:delta_force}
\end{equation}
where ${\bf F}^{(I)}_A$ is the force acting on nucleus $A$ in electronic state $I$. This difference directly probes the change in the potential-energy surface induced by the excitation, which in turn produces a small renormalization of the excitation energies in the absorption spectra.

\subsection{TDDFT and BSE}
\label{sec:tddft_bse}

Within the Tamm-Dancoff approximation (TDA) used here, a BSE excitation can be represented as a superposition
of electron-hole transitions,
\begin{equation}
|n\rangle_{\mathrm{BSE}} = \sum_{vc} A_{vc}^{n}\,|v\rightarrow c\rangle ,
\label{eq:bse_wf}
\end{equation}
where $n$ labels the BSE excited state, $v$ and $c$ denote occupied and unoccupied single-particle
states, respectively, and $A_{vc}^{n}$ is the amplitude of the corresponding electron-hole transition~\cite{Yu2024BSE}.

Equation~(\ref{eq:bse_wf}) has the same general form as the CIS expansion of Eq.~(\ref{eq:cis_wf}):
both describe the excitation through a superposition of single particle-hole transitions. This
correspondence allows us to compare the particle-hole composition of the BSE and CIS excited states.
The CIS amplitudes are expressed in the natural-orbital representation introduced above, whereas the
BSE amplitudes refer to the Kohn-Sham orbital basis. The $A_{vc}^{n}$ amplitudes also 
give access to the excited-state density in Eq.~\eqref{eq:delta_rho}, the transition
dipole moment, and the excitation-induced forces of Eq.~\eqref{eq:delta_force} through the 
corresponding analytical derivatives~\cite{Jin2026BSEForces}.

Within the TDA, TDDFT can be described in an analogous particle-hole representation.

\section{Computational Details}

The PySCF~\cite{PySCF} code is used to calculate the initial Hartree-Fock (HF) 
and CIS wave functions for the QMC calculations.
We employ the correlation-consistent effective core potential (ccECP) 
for hydrogen and the corresponding cc-pVDZ Gaussian basis set~\cite{Wang2019ccECP}. 
The Gaussian basis functions are periodized over the simulation cell. 
For each nuclear configuration, 
a singular-value decomposition (SVD) of the CIS amplitude matrix is used to 
obtain a compact spin-adapted
expansion~\cite{Martin2003,Mayer2007} of just $N/2$ CSFs
expressed in natural transition orbitals that represents the full CIS.

The QMC calculations are performed with CHAMP-EU~\cite{CHAMP}. 
For every nuclear configuration, the parameters of the 
ground- and excited-state wave functions are
optimized by state-specific energy minimization in 
VMC using the stochastic-reconfiguration 
method in a low-memory 
implementation~\cite{Sorella2007,Neuscamman2012}. 
Orthogonality to the ground state is imposed during 
the excited-state optimization through
a penalty constraint with weight $\lambda=1$~\cite{Pathak2021,Wheeler2024,Shepard2025}.

VMC forces are computed using a regularized force 
estimator~\cite{AttaccaliteSorella2008} with a node-cutoff parameter $\epsilon=0.1$ a.u.
In the DMC simulations~\cite{umrigar93} we use a time step of 0.05 a.u. and a population of 100 walkers. For quantities other than the energy, e.g. densities and structure factors, we report
the extrapolated estimator, twice the mixed minus the variational 
estimator~\cite{Foulkes2001}.

The TDDFT and BSE calculations are performed with WEST~\cite{west,Govoni2015,Yu2024BSE,Jin2023} starting from
Kohn-Sham calculations carried out with Quantum Espresso~\cite{Giannozzi2009,Giannozzi2017} and 
the PBE0 functional~\cite{Adamo1999}. We use
the SG15 optimized norm-conserving Vanderbilt pseudopotential for hydrogen~\cite{Schlipf2015} and a
plane-wave basis with a kinetic-energy cutoff of 200~Ry. 
The integrable singularity in the exact-exchange term is treated using the Gygi-Baldereschi 
scheme~\cite{Gygi1986} and the exact-exchange operator is evaluated using the adaptively compressed 
exchange representation~\cite{ACE}.

For the BSE calculations, dielectric screening is represented in the projective dielectric eigenpotential
basis. Quasiparticle corrections are evaluated within $G_0W_0$ starting from PBE0 and included
in the BSE Hamiltonian. The resulting $G_0W_0$-BSE@PBE0 calculations are used for the vertical excitation
energies, BSE amplitudes, excited-state density differences, and transition dipole moments. 
Excited-state forces are instead calculated at the
BSE@PBE0 level, without the $G_0W_0$ quasiparticle corrections, using the analytical-force implementation
of Ref.~\cite{Jin2026BSEForces}. 

The TDDFT calculations employ the TDA and the same PBE0 reference,
pseudopotentials, cutoffs, and simulation cells as the BSE calculations. 
Excited-state forces
are obtained from the analytical-force implementation for hybrid TDDFT~\cite{Jin2023}, 
in which the Lagrangian formulation and the Z-vector equation are used 
to evaluate the derivatives of the excitation energy.

\section{Results}
\label{sec:results}

\subsection{Excitation energies and particle-hole character}
\label{sec:results_gap}

\begin{figure}[t]
\includegraphics[width=\columnwidth]{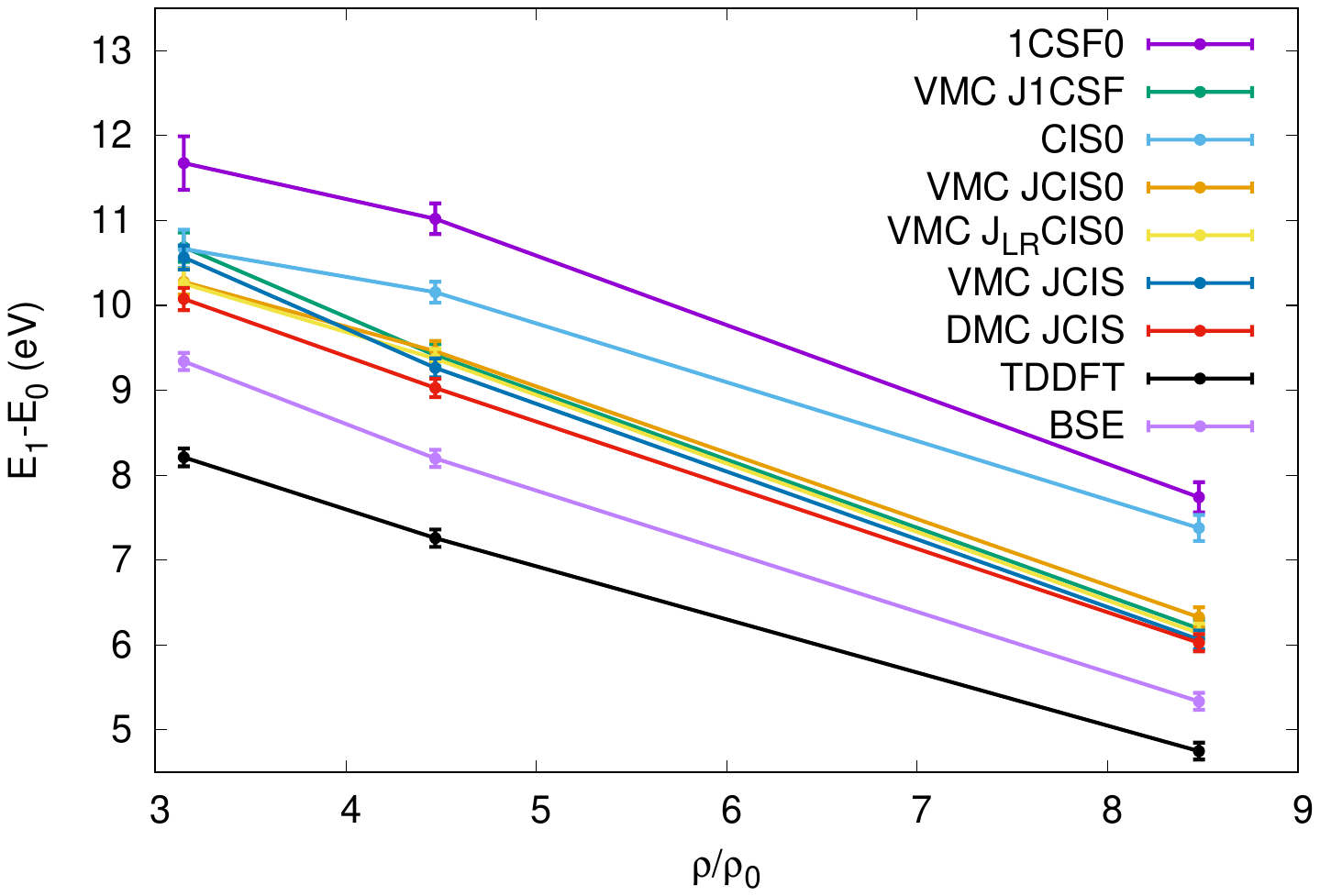}
\caption{Neutral excitation gap of phase-I molecular hydrogen as a function of
compression $\rho/\rho_0$. For each density, the results are averaged over
the nuclear configurations of the 96-atom simulation cell. Error bars
indicate the statistical uncertainty associated with the configurational
average; for the QMC results, the statistical errors on excitation
energies are much smaller.
}
\label{fig:gap}
\end{figure}

Figure~\ref{fig:gap} shows the neutral excitation gap as a function of the density.
The various QMC results are rather close to each other, with the spread decreasing
at higher density and DMC with {\em JCIS} wave functions consistently giving the 
lowest QMC values.
In particular, the VMC {\em J1CSF} and VMC {\em JCIS} gaps are very similar despite the
different flexibility of their excited-state wave functions. The similarity of
the VMC {\em J1CSF} and VMC {\em JCIS} gaps suggests that the excitation energy is only weakly
affected by the explicit CIS expansion once the Jastrow factor and orbitals are
optimized. The $G_0W_0$-BSE@PBE0 (denoted simply as BSE) gaps lie below the 
DMC {\em JCIS} values by approximately $0.7$--$0.8$ eV, while the TDDFT gaps are even lower.

\begin{figure}[t]
\centering
\includegraphics[width=\columnwidth]{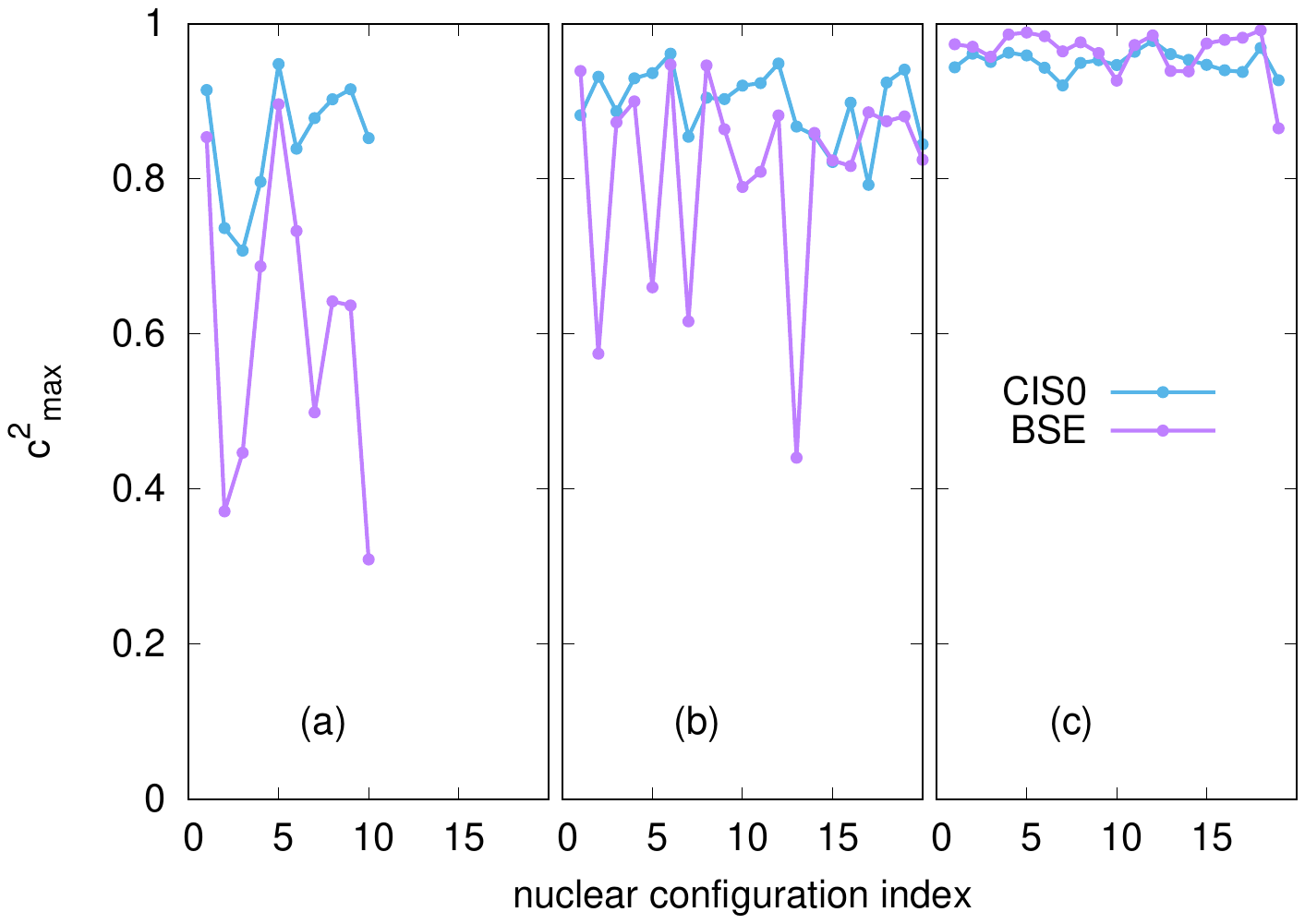}
\caption{Largest squared coefficient, $c_{\rm max}^2$, in the {\em CIS0} expansion and
corresponding largest squared BSE transition amplitude 
for each nuclear configuration at compressions $\rho/\rho_0=3.15$, $4.47$, and $8.48$ [panels (a)--(c)].
Both {\em CIS0} and BSE show an evolution toward an excitation dominated by a
single particle-hole pair, $c_{\rm max}^2\rightarrow 1$, with increasing
density.
}
\label{fig:cmax}
\end{figure}

The relative behavior of the bare {\em CIS0} gap changes more strongly with density. At the highest
density, the {\em CIS0} gap is close to the HF value, while it moves closer 
to the correlated QMC results at lower density. This larger effect of the {\em CIS0}
expansion at low density is accompanied by a stronger multiconfigurational character 
of the excited state, with contributions from several particle-hole pairs. Figure~\ref{fig:cmax}
shows the largest squared coefficient, $c_{\rm max}^2$, in the {\em CIS0} expansion
for each nuclear configuration, together with the corresponding largest
squared BSE transition amplitude, $|A_{vc}^{1}|^2$ (Eq.~\ref{eq:bse_wf}).
At the lowest density, several configurations
have values appreciably below one, indicating significant contributions from
more than one particle-hole transition. With increasing density,
$c_{\rm max}^2$ approaches one for both {\em CIS0} and BSE, and the excitation becomes
increasingly dominated by a single particle-hole pair.
We note that, at low density, the {\em CIS0} expansion is systematically more compact 
than the BSE representation, consistent with the use of natural orbitals 
in the {\em CIS0} wave function as opposed to canonical orbitals in BSE.

Therefore, even though the VMC {\em J1CSF} and VMC {\em JCIS} excitation energies are very similar,
the particle-hole analysis reveals differences in the character of the
excited state, as becomes clear from its spatial properties discussed below.

\subsection{Exciton localization}
\label{sec:results_localization}

\begin{figure}[t]
\centering
\includegraphics[width=\columnwidth]{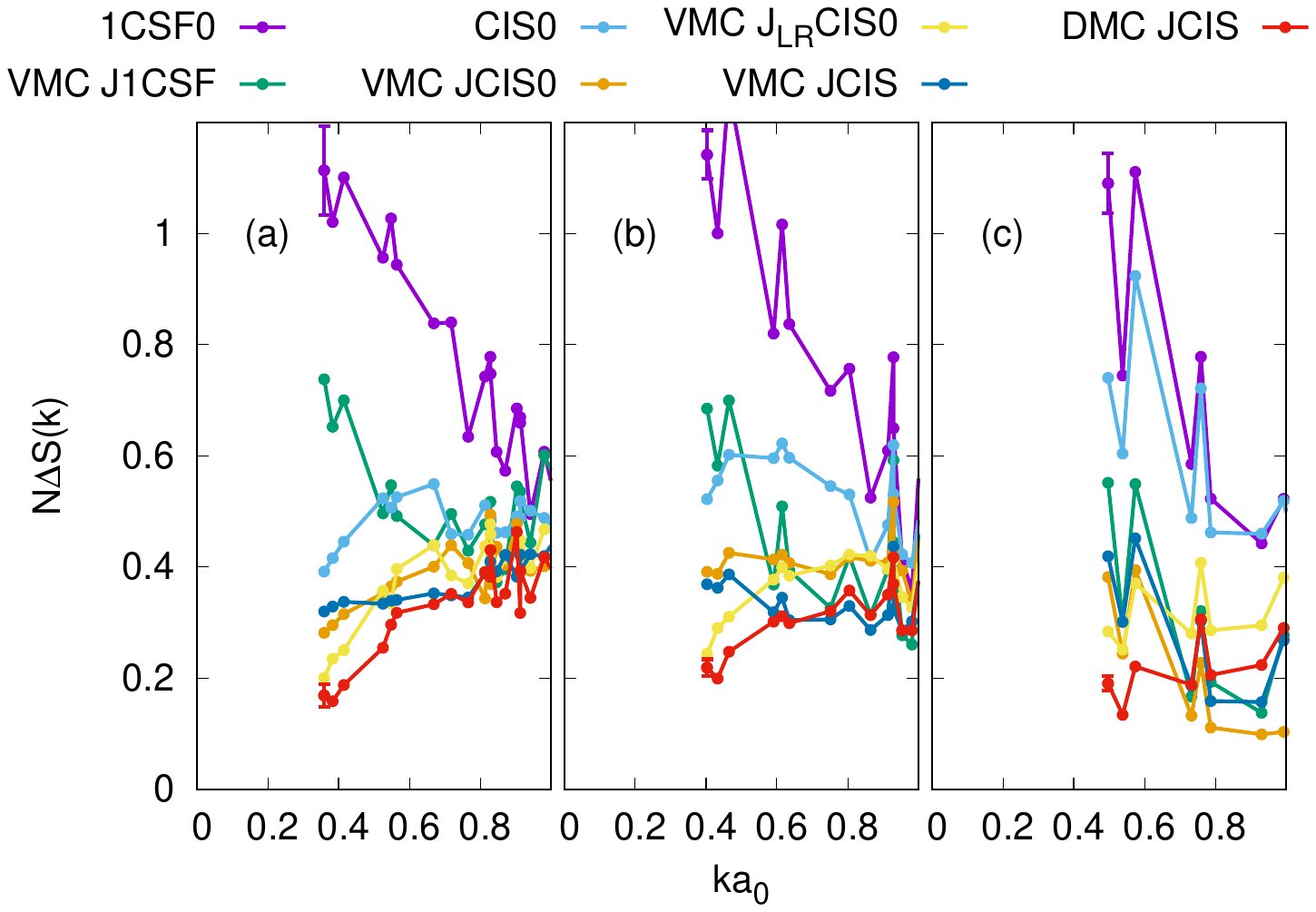}
\caption{Difference $N\Delta S(k)$ between the static structure factors of the
excited and ground states, averaged over the nuclear configurations, at
compressions $\rho/\rho_0=3.15$, $4.47$, and $8.48$
[panels (a)--(c)]. Statistical errors, dominated by the configurational
average, are shown for {\em 1CSF0} and DMC {\em JCIS} at the smallest wave vector.
}
\label{fig:deltaS}
\end{figure}

Figure~\ref{fig:deltaS} shows the small-$k$ behavior of the difference
$\Delta S(k)$ between the static structure factors of the excited and ground
states, averaged over the nuclear configurations. The curves are not
smooth, 
due to finite-size reciprocal-lattice shell effects,
but clear trends emerge among the different
wave functions. As discussed in Sec.~\ref{sec:exciton_characterization}, a
vanishing $\Delta S(k)$ in the $k\rightarrow 0$ limit is consistent with an
excitation localized within the simulation cell.

At the lowest density, {\em 1CSF0} gives the largest $\Delta~S(k)$ in the
long-wavelength limit.
The VMC {\em J1CSF} wave function reduces it substantially but the curve
still increases as $k$ decreases. Therefore, the inclusion
of the Jastrow factor and state-specific orbital optimization does not by itself
produce the localization obtained when the CIS expansion is introduced. 
For {\em CIS0}, $\Delta S(k)$ instead decreases toward small $k$ in a 
manner consistent with a vanishing limit for $k\rightarrow 0$. 
Adding explicit correlation through the
Jastrow factor lowers $\Delta S(k)$ further, both with (VMC {\em JCIS}) 
and without (VMC {\em JCIS0}) reoptimization of the orbitals and expansion 
coefficients. 
The reduction is stronger when the long-range contribution to the
electron--electron Jastrow factor is included (VMC {\em J}$_{LR}${\em CIS0}):
the correct long-range behavior of the RPA term, apparently not recovered
by VMC energy minimization, further promotes localization. 
A similar effect is provided by imaginary time evolution, with DMC {\em JCIS}
reaching the smallest values of $\Delta~S(k)$.

The same qualitative behavior is found at the two higher densities, but the
effect of the CIS expansion becomes progressively weaker. This trend is
consistent with the evolution of $c_{\rm max}^2$ in Fig.~\ref{fig:cmax}: as
the excitation becomes increasingly dominated by a single particle-hole pair,
the difference between the single-CSF and CIS descriptions is reduced.

\begin{figure*}[h]

\hspace{-1.0cm}
\begin{subfigure}{0.17\textwidth}
\captionsetup{labelformat=empty}
\caption{{\em 1CSF0}}
\includegraphics[width=3.0cm,height=3.0cm]{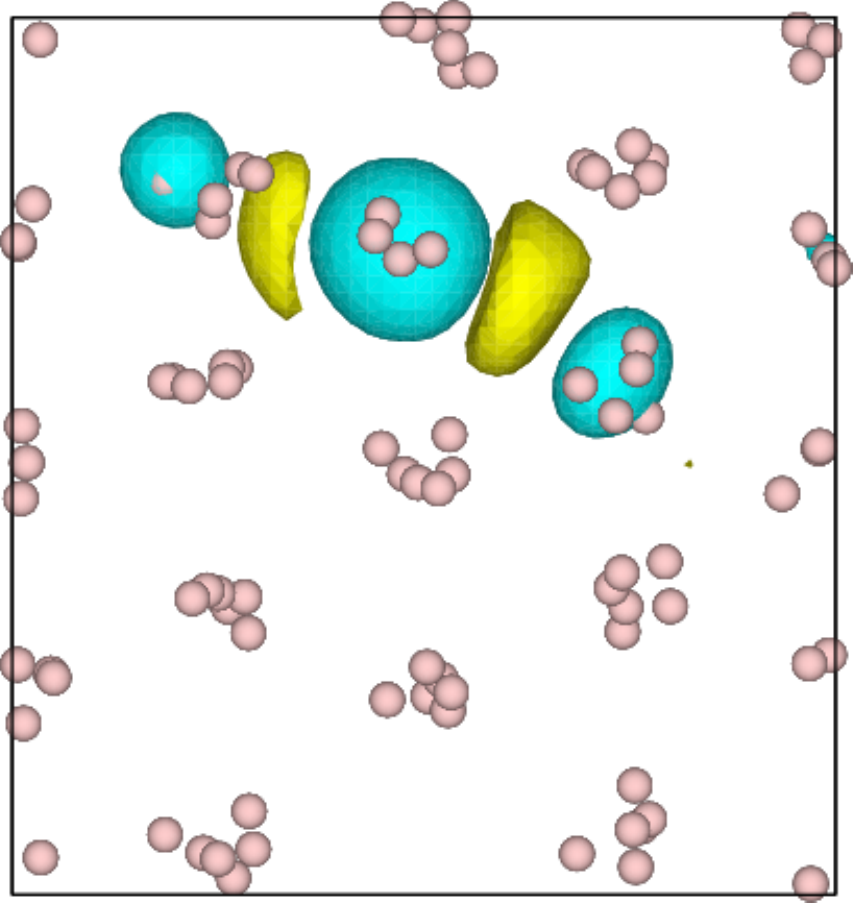}
\label{hf_1}
\end{subfigure}
\begin{subfigure}{0.17\textwidth}
\captionsetup{labelformat=empty}
\caption{DMC {\em J1CSF}}
\includegraphics[width=3.0cm,height=3.0cm]{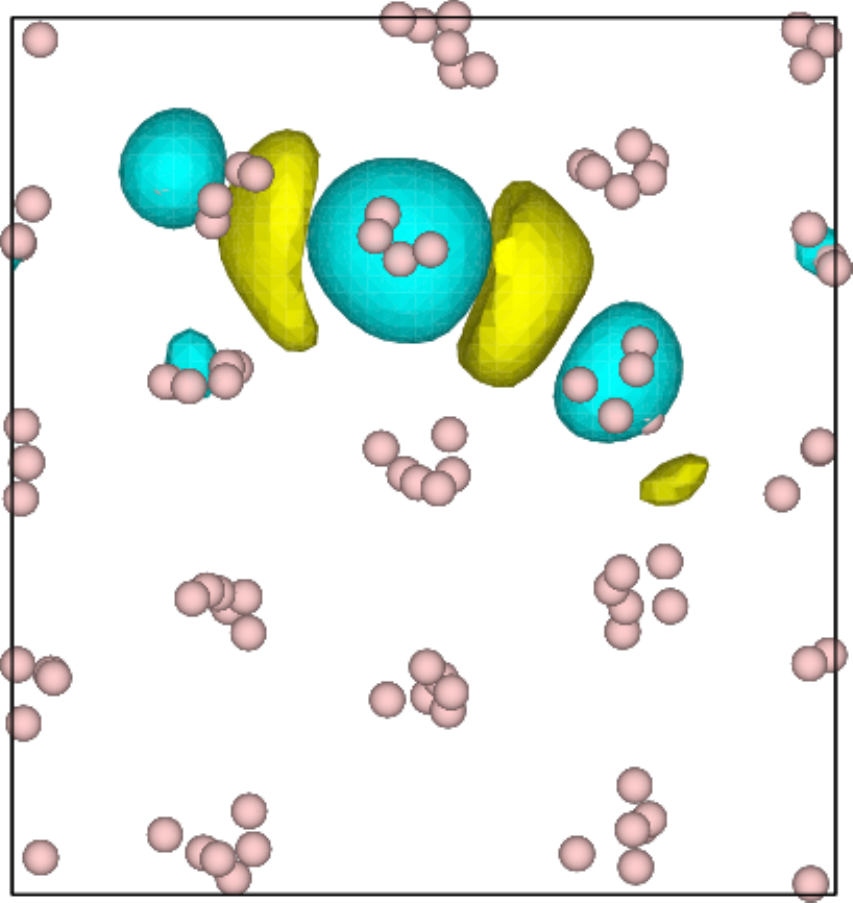}
\label{vmc_1}
\end{subfigure}
\begin{subfigure}{0.17\textwidth}
\captionsetup{labelformat=empty}
\caption{TDDFT}
\includegraphics[width=3.0cm,height=3.0cm]{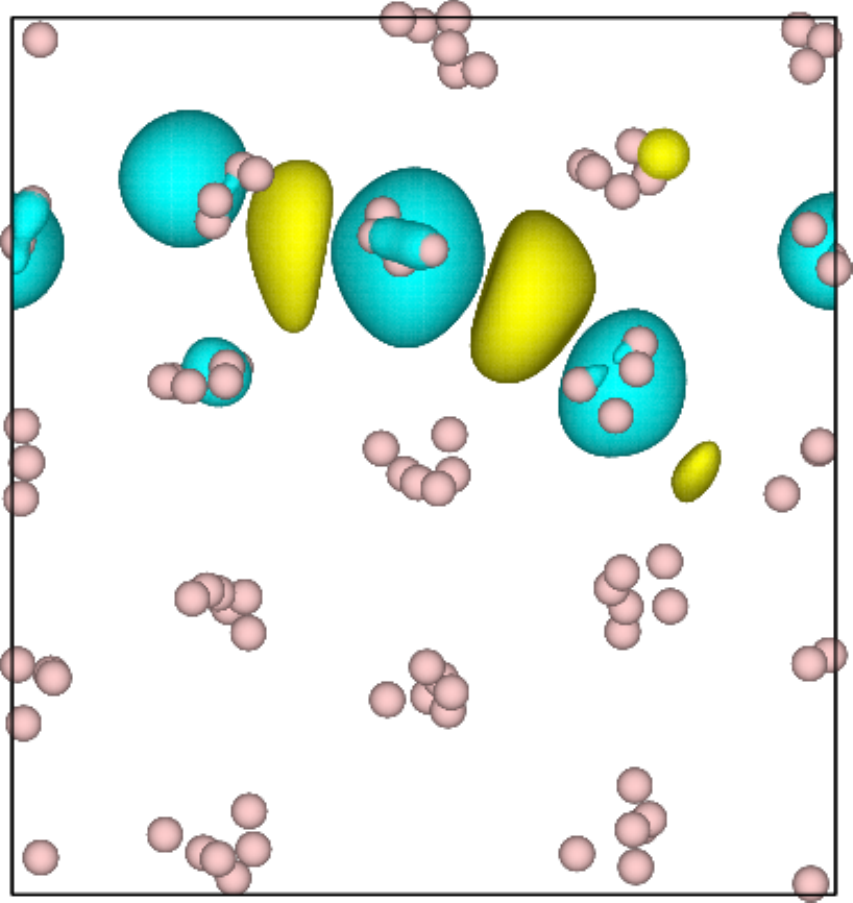}
\label{pbe0_1}
\end{subfigure}
\begin{subfigure}{0.17\textwidth}
\captionsetup{labelformat=empty}
\caption{BSE}
\includegraphics[width=3.0cm,height=3.0cm]{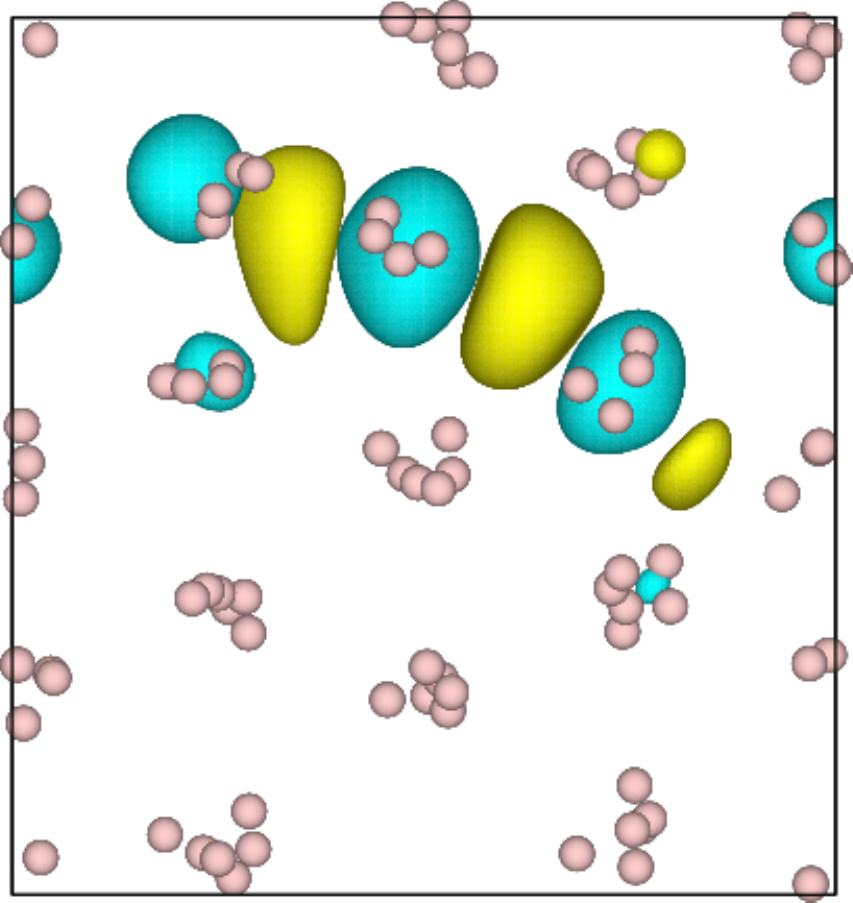}
\label{bse_1}
\end{subfigure}
\begin{subfigure}{0.17\textwidth}
\captionsetup{labelformat=empty}
\caption{{\em CIS0}}
\includegraphics[width=3.0cm,height=3.0cm]{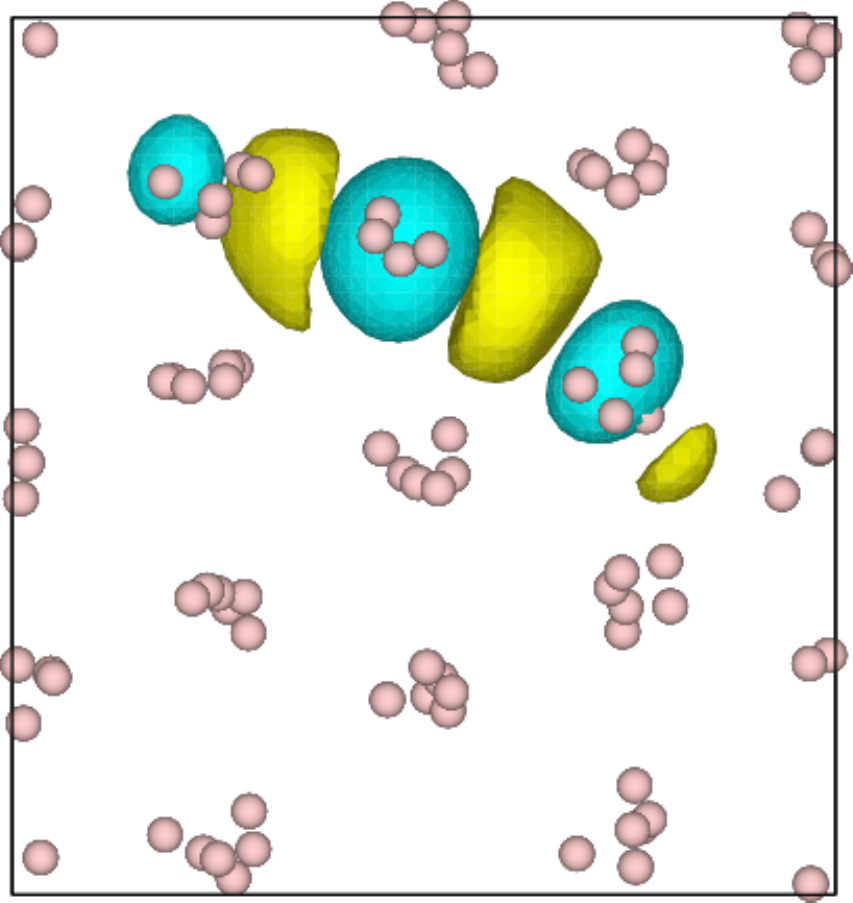}
\label{cis_1}
\end{subfigure}
\begin{subfigure}{0.17\textwidth}
\captionsetup{labelformat=empty}
\caption{DMC {\em JCIS}}
\begin{tikzpicture}
\node at (0,0){\includegraphics[width=3.0cm,height=3.0cm]{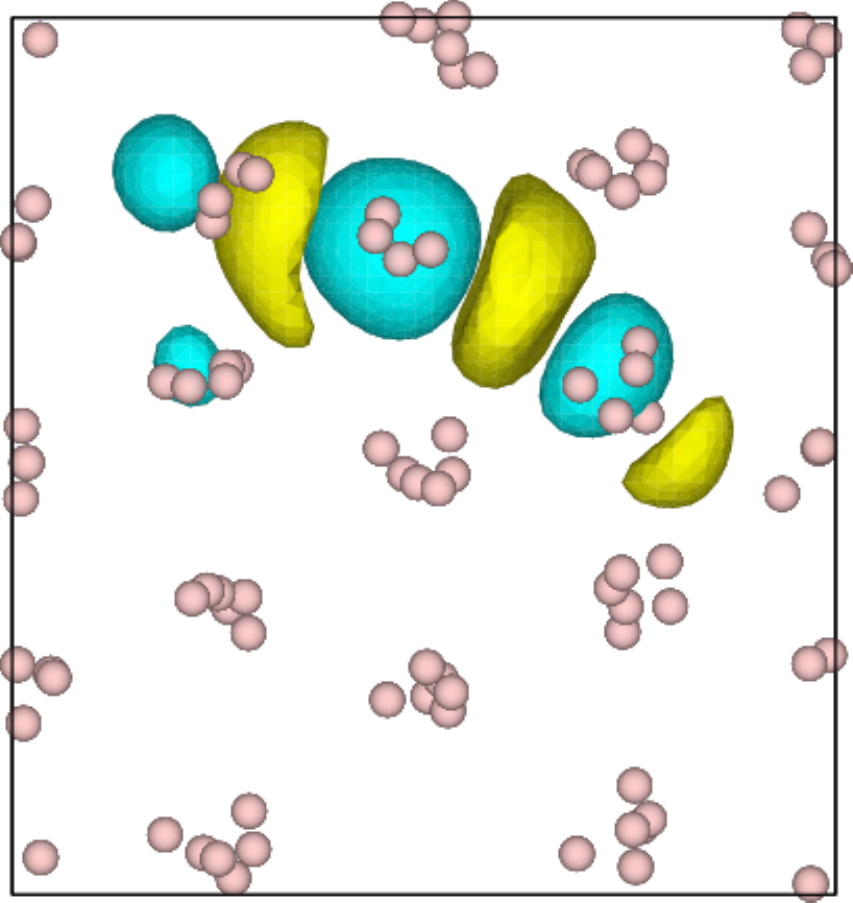}};
\draw[] node at (2,0){\bf a1};
\end{tikzpicture}
\label{dmc_1}
\end{subfigure}

\hspace{-1.0cm}
\begin{subfigure}{0.17\textwidth}
\includegraphics[width=3.0cm,height=3.0cm]{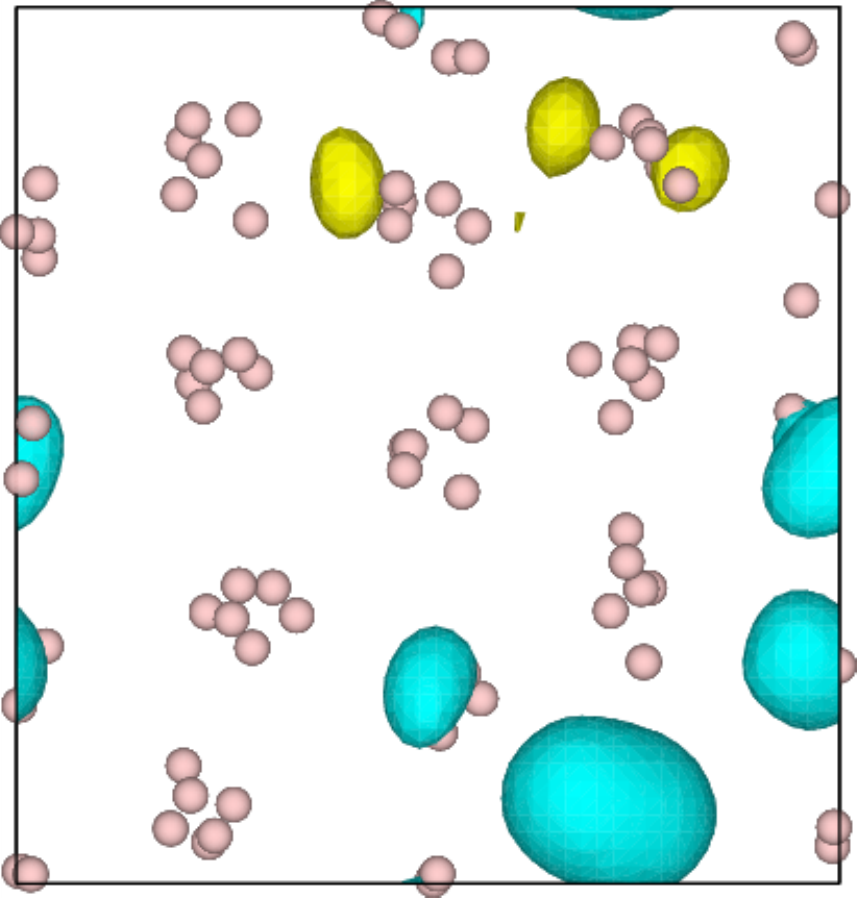}
\label{hf_2}
\end{subfigure}
\begin{subfigure}{0.17\textwidth}
\includegraphics[width=3.0cm,height=3.0cm]{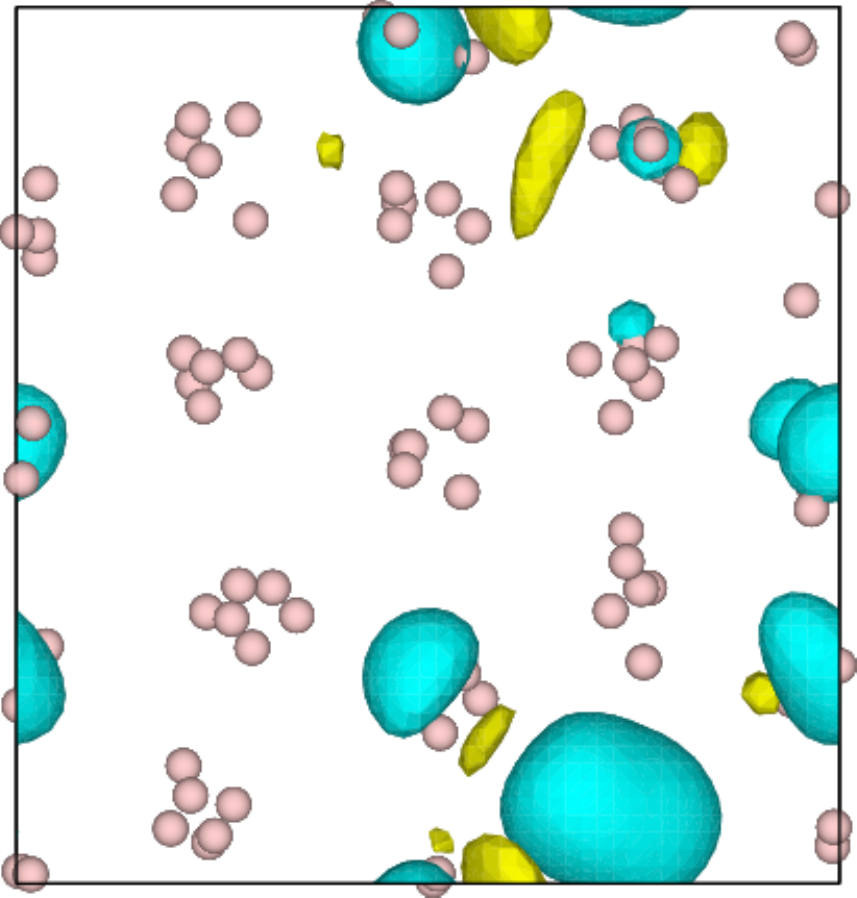}
\label{vmc_2}
\end{subfigure}
\begin{subfigure}{0.17\textwidth}
\includegraphics[width=3.0cm,height=3.0cm]{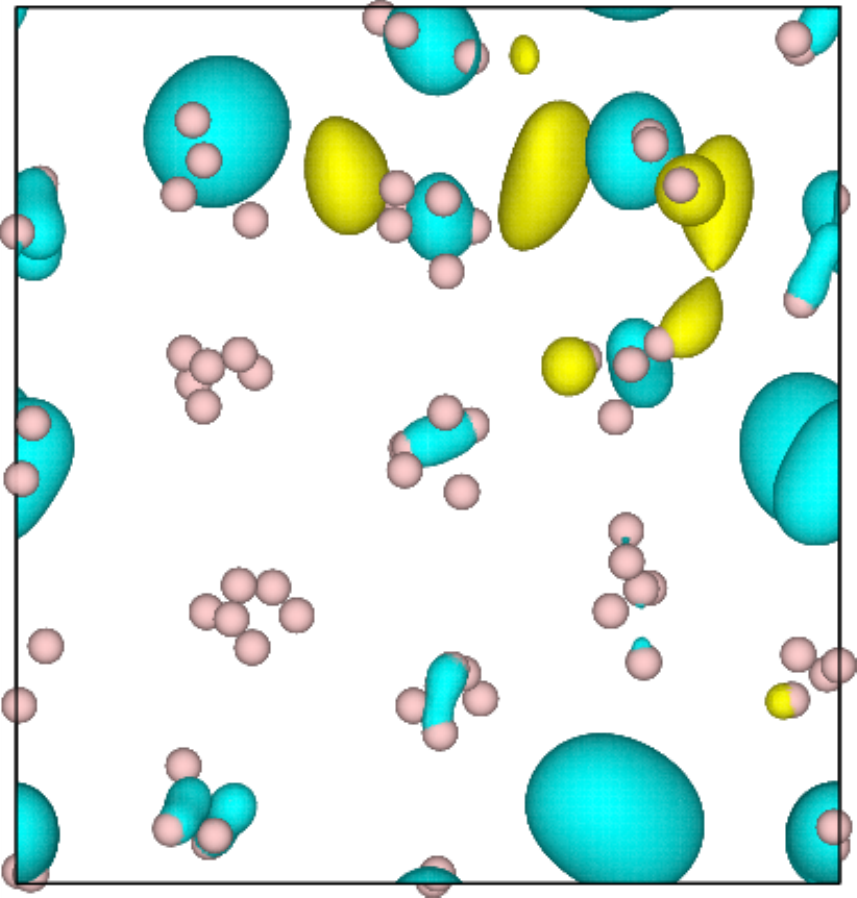}
\label{pbe0_2}
\end{subfigure}
\begin{subfigure}{0.17\textwidth}
\includegraphics[width=3.0cm,height=3.0cm]{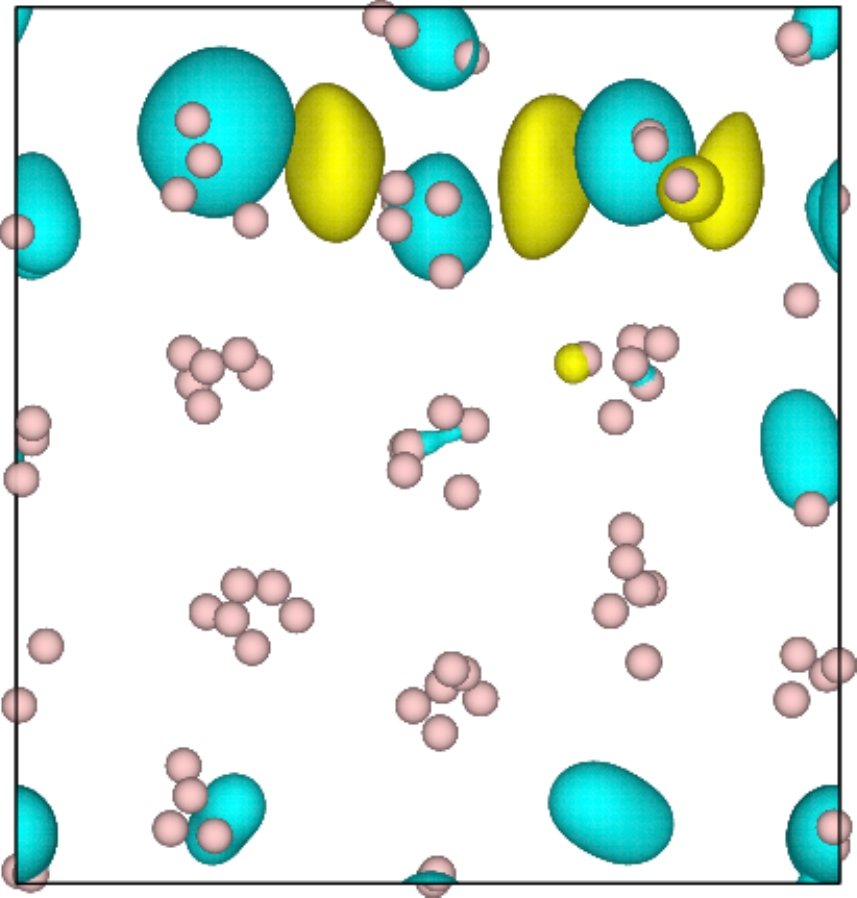}
\label{bse_2}
\end{subfigure}
\begin{subfigure}{0.17\textwidth}
\includegraphics[width=3.0cm,height=3.0cm]{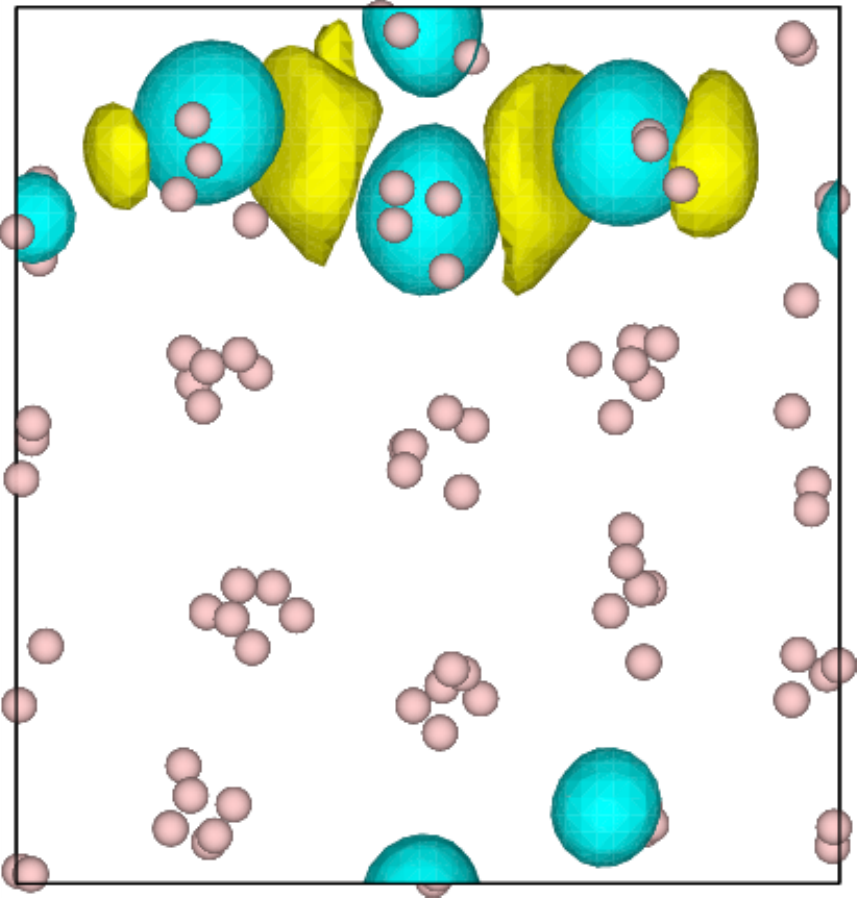}
\label{cis_2}
\end{subfigure}
\begin{subfigure}{0.17\textwidth}
\begin{tikzpicture}
\node at (0,0){\includegraphics[width=3.0cm,height=3.0cm]{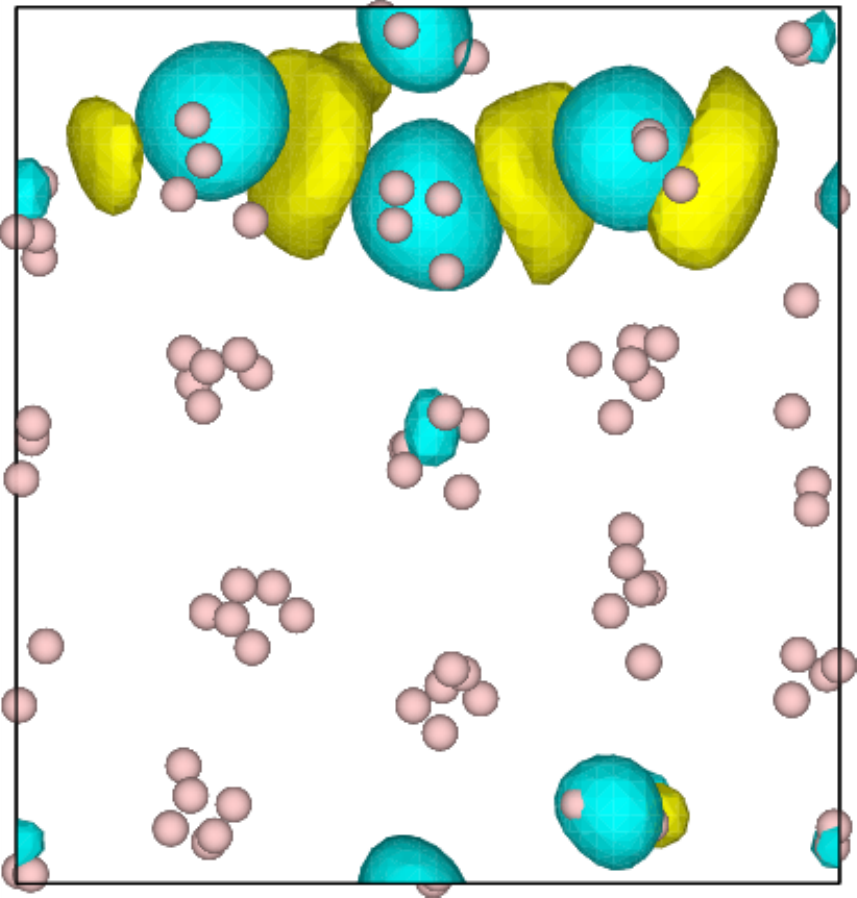}};
\draw[] node at (2,0){\bf a2};
\end{tikzpicture}
\label{dmc_2}
\end{subfigure}

\hspace{-1.0cm}
\begin{subfigure}{0.17\textwidth}
\includegraphics[width=3.0cm,height=3.0cm]{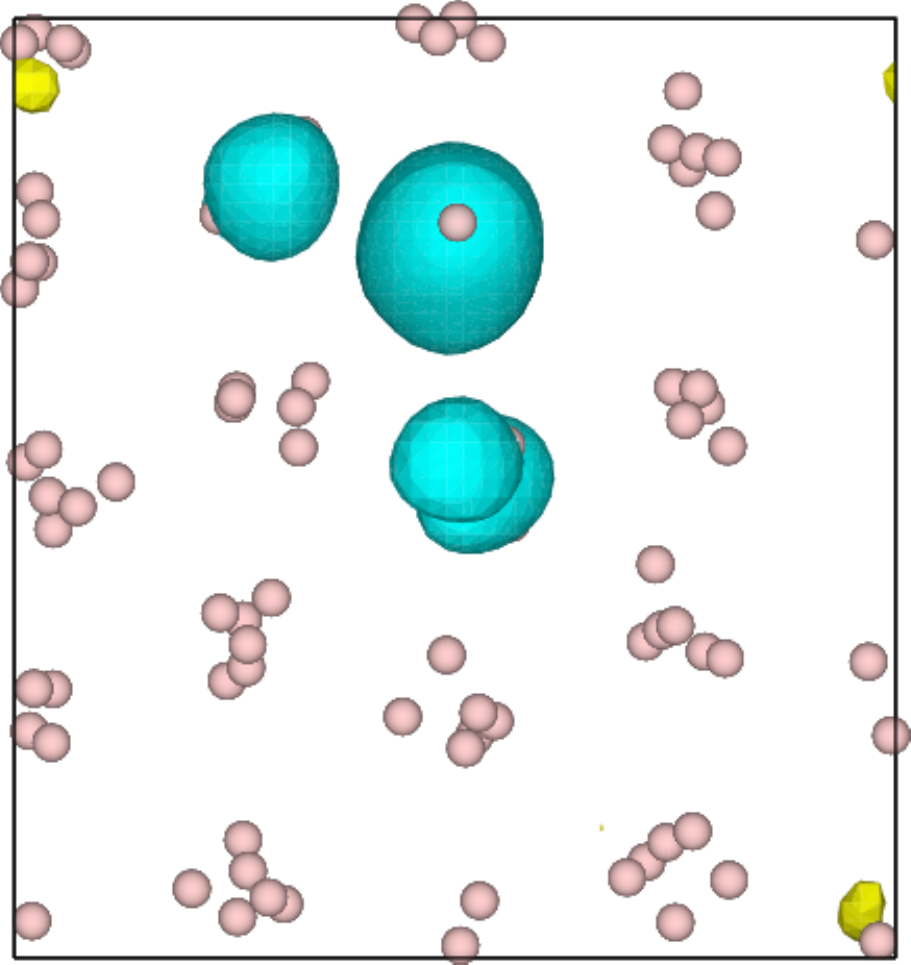}
\label{hf_8}
\end{subfigure}
\begin{subfigure}{0.17\textwidth}
\includegraphics[width=3.0cm,height=3.0cm]{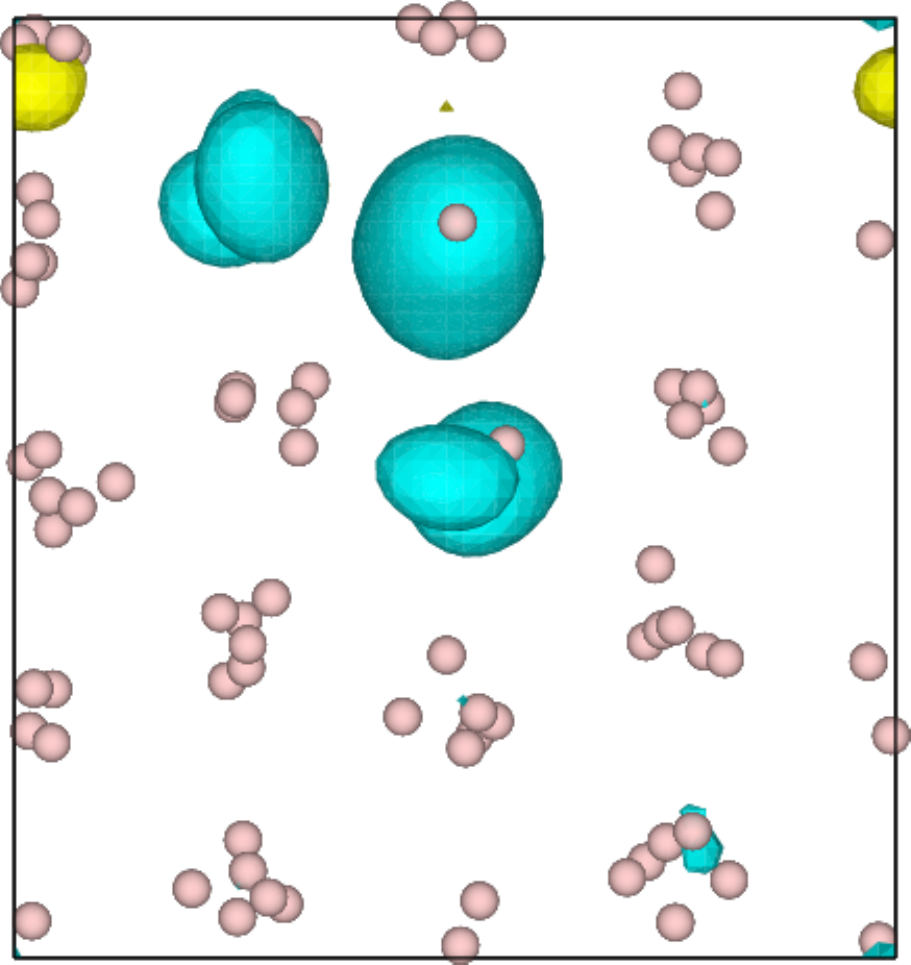}
\label{vmc_8}
\end{subfigure}
\begin{subfigure}{0.17\textwidth}
\includegraphics[width=3.0cm,height=3.0cm]{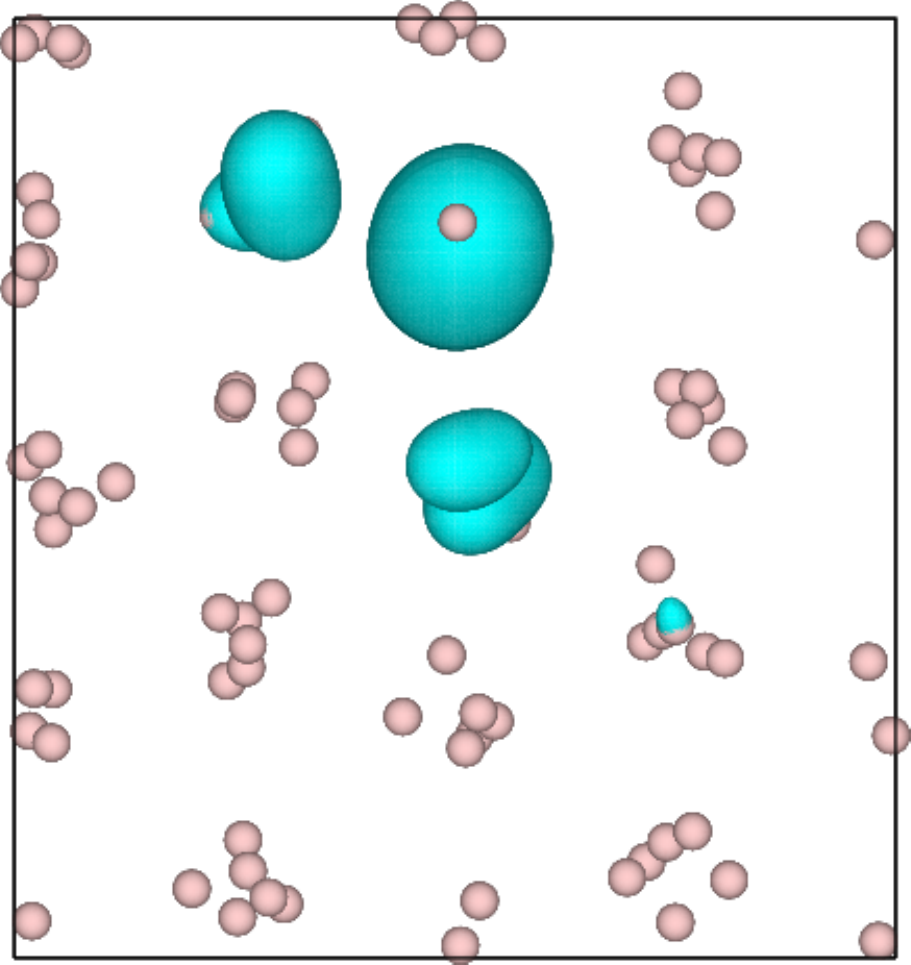}
\label{pbe0_8}
\end{subfigure}
\begin{subfigure}{0.17\textwidth}
\includegraphics[width=3.0cm,height=3.0cm]{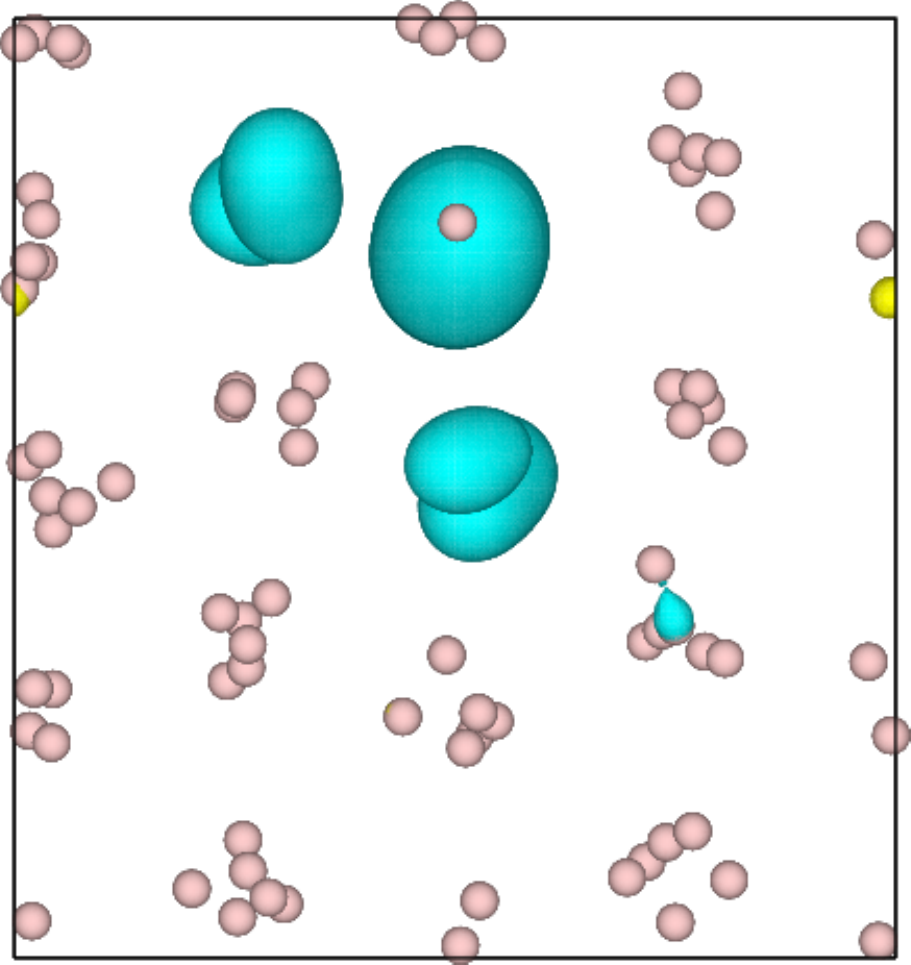}
\label{bse_8}
\end{subfigure}
\begin{subfigure}{0.17\textwidth}
\includegraphics[width=3.0cm,height=3.0cm]{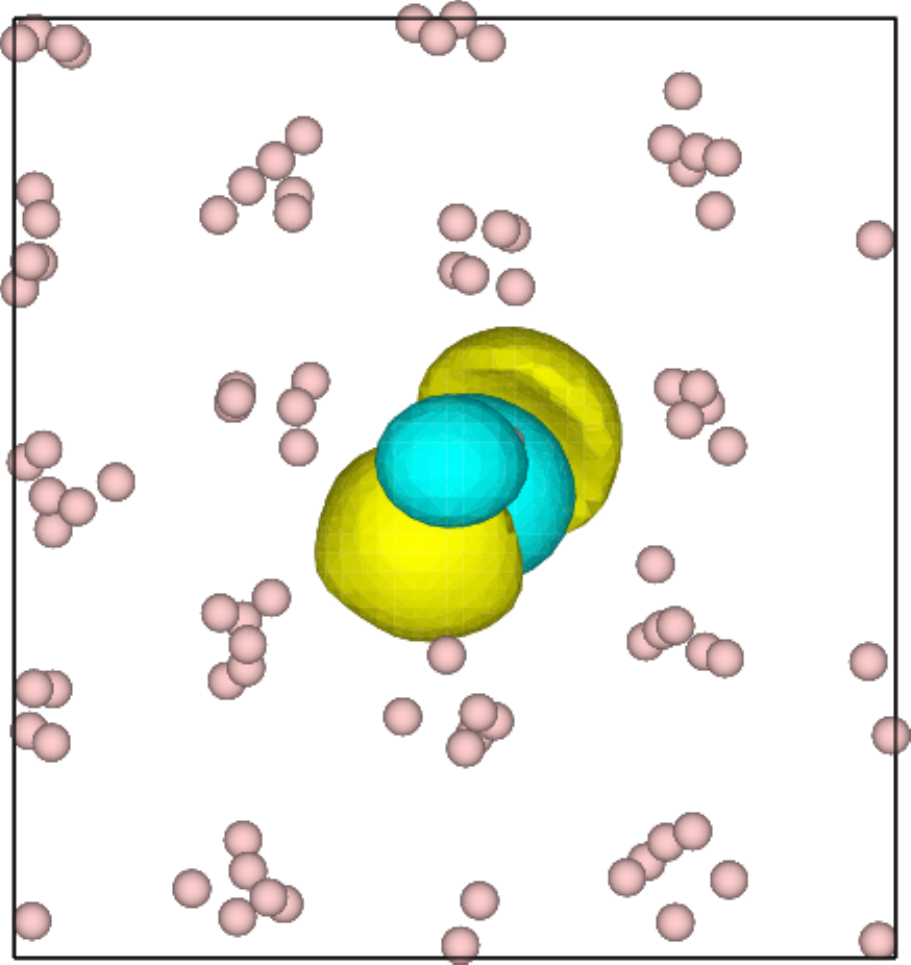}
\label{cis_8}
\end{subfigure}
\begin{subfigure}{0.17\textwidth}
\begin{tikzpicture}
\node at (0,0){\includegraphics[width=3.0cm,height=3.0cm]{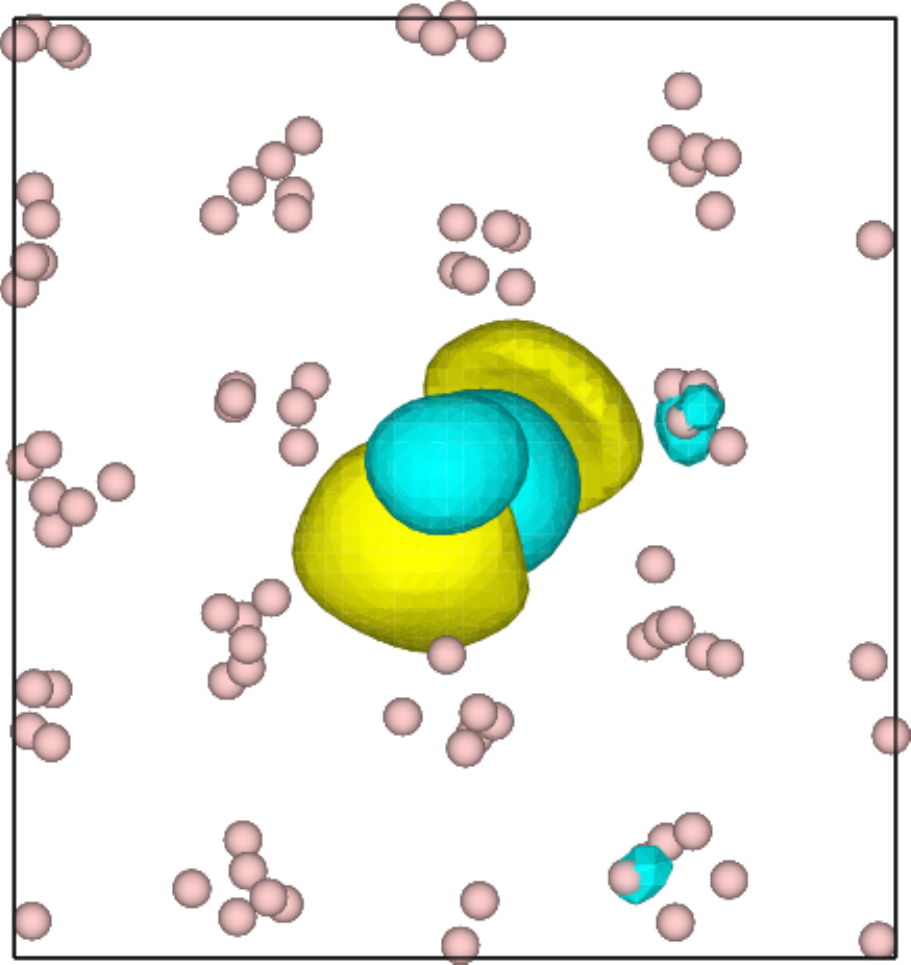}};
\draw[] node at (2,0){\bf a8};
\end{tikzpicture}
\label{dmc_8}
\end{subfigure}

\caption{Excitation-induced density difference, $\Delta\rho({\bf r})$, for selected
nuclear configurations at $\rho/\rho_0=3.15$. Configurations {\bf a1}, {\bf a2}, and {\bf a8}
illustrate different responses to the electronic structure treatment:
localization already present at the HF level, progressive localization from
HF through TDDFT and BSE toward the QMC result, and localization emerging
predominantly with the CIS-based QMC wave functions, respectively.
Positive (yellow) and negative (blue) isosurfaces indicate accumulation
and depletion of electronic density. The level of the isosurfaces is the
default of the VESTA visualization program~\cite{vesta}, $\pm\left[\langle|\Delta\rho|\rangle+2\sigma(|\Delta\rho|)\right]$, 
for all methods and configurations. Smallness or absence of yellow patches indicates a considerably smeared 
particle in the excitonic pair.
}
\label{fig:delta_rho}
\end{figure*}

A theoretical estimate of the leading contribution to the finite size 
effect $\Delta_n$ on the optical gap, which bypasses explicit extrapolation with different system sizes, 
is given by $\Delta_n=N\Delta_S(k\to 0)|v_M|$~\cite{Gorelov_2023}, 
where $v_M$ is the Madelung constant~\cite{Yang_2020}.
In Ref.~\cite{Gorelov_2024}, QMC simulations were based on a single determinant
trial wave function, resulting in a non-vanishing $N\Delta S(k \to 0)$ similar to our DMC {\em J1CSF}
results. This behavior, indicating
the absence of localization, was dismissed as an artifact of
the use of extended Bloch orbitals
in the determinant of the trial wave function.
In the absence of a reliable value for $N\Delta S(k \to 0)$,
localization at the two lowest densities studied here, and lack thereof for the higher density,
was instead taken for granted based on visual inspection of the BSE wave function of the exciton.
Conclusions on the size effect were then reached indirectly, based on estimates
of the localization length from the dielectric constant and the particle-hole band mass~\cite{Gorelov_2024}.

Using CIS wave functions, we are able to directly assess finite size corrections
in terms of the structure factor.
From the only non-vanishing value $N \Delta S(k \to 0) \approx 0.1$ at the highest density
(see DMC {\em JCIS} in Fig.~\ref{fig:deltaS}), 
we infer that the corresponding excitation gap will be increased by about $0.6$eV in the bulk limit. 
This value, in agreement with the conclusions reached in Ref.~\cite{Gorelov_2024}, is now obtained
in a coherent framework, without invoking BSE input.

The configurationally averaged structure factor does not show how the degree of
localization varies among individual nuclear configurations. We therefore examine
the difference in real space $\Delta\rho({\bf r})$ of the electronic densities of the excited and the ground state, 
which also offers a straight and detailed comparison between various methods.
Figure~\ref{fig:delta_rho} shows three representative configurations 
{\bf a1}, {\bf a2} and {\bf a8} at the
lowest density, $\rho/\rho_0=3.15$; results for all configurations are reported
in the Supplemental Material~\cite{suppmat}.  

In configuration {\bf a1}, the excitation is already localized at the {\em 1CSF0} level. The density is depleted
mainly around a small group of neighboring molecules and accumulated in the
surrounding interstitial regions, with only minor differences among the
different electronic structure methods. In this case, disorder in the nuclear 
positions is sufficient to induce localization.
In configuration {\bf a2}, localization develops progressively with the level of theory.
The {\em 1CSF0} and DMC {\em J1CSF} density differences remain extended over the
simulation cell, while TDDFT produces a more localized pattern and BSE better
resembles the {\em CIS0} and DMC {\em JCIS} results.
In configuration {\bf a8}, we observe a different situation. Here, {\em 1CSF0}, DMC {\em J1CSF}, TDDFT, and
BSE show density changes spread over much of the simulation cell, while {\em CIS0} and DMC {\em JCIS} produce
a more localized excitation, with depletion concentrated around two neighboring 
molecules and accumulation in the nearby interstitial regions.  The additional 
flexibility provided by the CIS expansion, both with and without Jastrow factors or
imaginary time projection, is therefore essential for this configuration.

The three examples show that the role of the electronic structure treatment
depends strongly on the local nuclear environment. Disorder can already localize
the excitation at the mean-field level, as in {\bf a1}, or localization can develop
progressively from HF through TDDFT and BSE toward QMC, as in {\bf a2}. In other
configurations, such as {\bf a8}, the localized character only emerges upon inclusion
of the CIS expansion. The real-space density differences therefore support the
conclusion drawn from $\Delta S(k)$ that optimization of a single-CSF wave
function in general fails to reproduce the spatial character obtained with the
multideterminant excited state.

With increasing density, the differences among the electronic structure methods
become less pronounced (see Supplemental Material~\cite{suppmat}). At the intermediate density, the same three 
types of behavior are still found among different nuclear configurations. At the 
highest density, the density changes are more similar 
across the electronic structure methods, consistent with the weaker
multiconfigurational character of the excitation and with the reduced effect of
the CIS expansion on $\Delta S(k)$.

\subsection{Optical and nuclear response}
\label{sec:results_tdm}

\begin{figure}[t]
\centering
\includegraphics[width=\columnwidth]{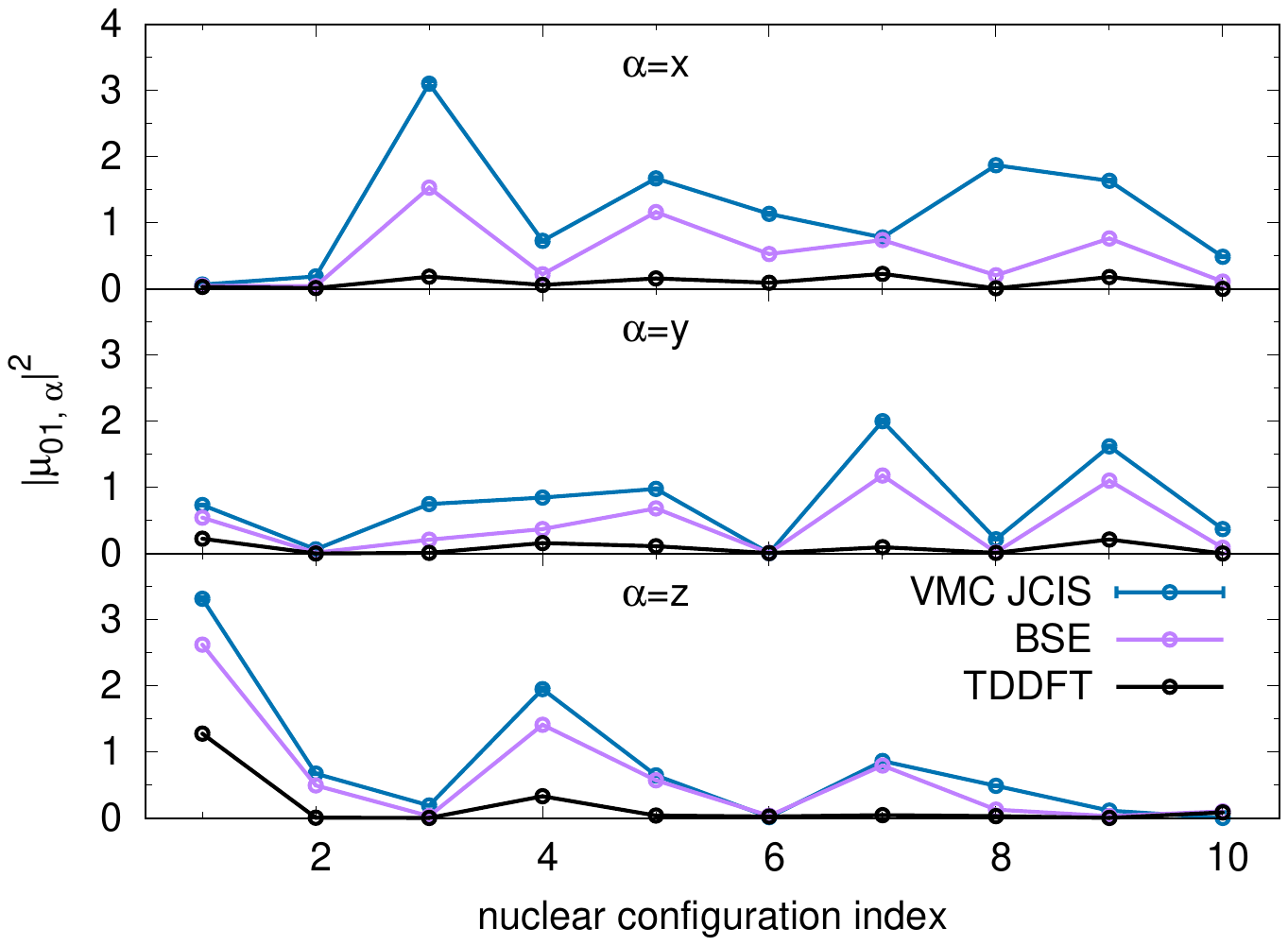}
\caption{Square modulus of the Cartesian components of the ground-to-excited-state transition dipole
moment, in atomic units, for the ten nuclear configurations at
$\rho/\rho_0=3.15$, obtained with VMC {\em JCIS}, BSE, and TDDFT.
The statistical errors on the VMC results are smaller than the symbol size.
}
\label{fig:tdm}
\end{figure}

We next examine the optical character of the excitation through the transition
dipole moment. Figure~\ref{fig:tdm} shows the square of its Cartesian 
components for the ten nuclear configurations at the lowest density, 
$\rho/\rho_0=3.15$. The values
vary substantially from one configuration to another both in
magnitude and orientation.

BSE follows the QMC results reasonably well, with the notable exception of
the $x$ component for configuration {\bf a8}, in correspondence with the difference observed
for $\Delta\rho({\bf r})$, while
TDDFT shows larger deviations and generally weaker variation among the
configurations. More results on the TDM are collected in the Supplemental Material~\cite{suppmat}.

\begin{figure}[t]
\centering
\includegraphics[width=\columnwidth]{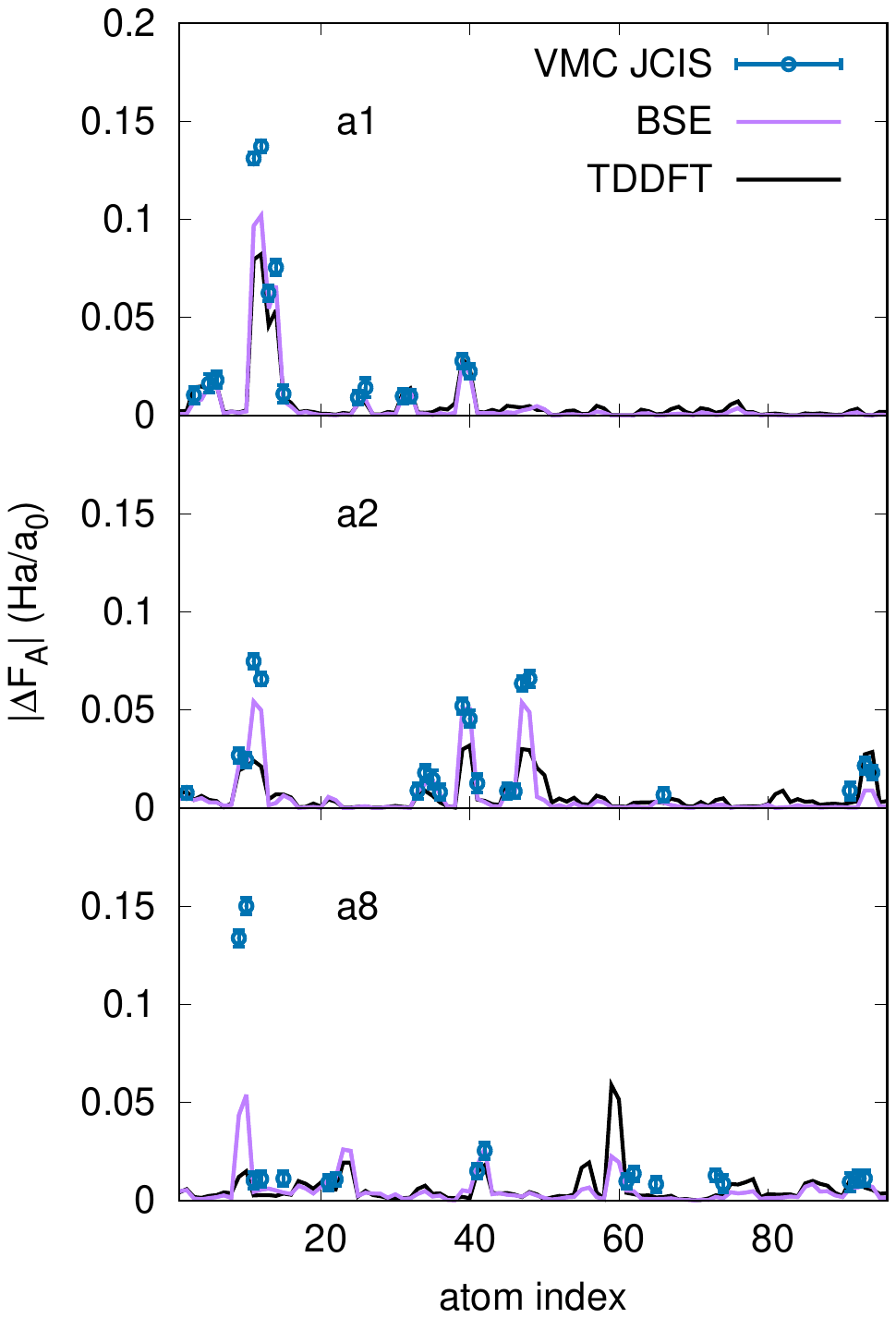}
\caption{Magnitude of the excitation-induced force,
$|\Delta{\bf F}_A|$, on each atom for configurations {\bf a1}, {\bf a2}, and {\bf a8} at
$\rho/\rho_0=3.15$, obtained with VMC {\em JCIS}, BSE@PBE0, and TDDFT. The same
configurations are shown in Fig.~\ref{fig:delta_rho}.
}
\label{fig:deltaF}
\end{figure}

Figure~\ref{fig:deltaF} shows the magnitude of the excitation-induced force,
$|\Delta{\bf F}_A|$, for the same three configurations considered in
Fig.~\ref{fig:delta_rho}. 
$\Delta{\bf F}_A$ conveys an even more detailed information,
as it refers to individual atoms within a given configuration.
For configuration {\bf a1}, QMC (VMC {\em JCIS}), BSE@PBE0 (denoted BSE below), 
and TDDFT show the strongest force changes
between ground and excited states on a similar set of atoms, with some differences 
in magnitude.  For {\bf a2}, the force magnitudes differ more among the methods,
paralleling the more pronounced density changes between TDDFT, BSE and QMC in
Fig.~\ref{fig:delta_rho}.

The largest discrepancies are found for {\bf a8}. QMC gives large force changes on a
small number of atoms, while the corresponding BSE and TDDFT forces are much
smaller and include additional peaks on different atoms. 
This is indeed the configuration for which the {\em CIS0} and DMC {\em JCIS} $\Delta\rho({\bf r})$
are substantially more localized than their BSE and TDDFT counterparts.
The differences in the spatial description of the excitation are therefore directly
reflected in the excited-state forces acting on the nuclei.

\section{Conclusions}
\label{sec:conclusions}

QMC excitation energies and spatial properties respond differently to the
choice and optimization of the excited-state wave function. Once the Jastrow
factor and orbitals of the single-CSF wave function are optimized, its
excitation energy is very close to that obtained with the CIS expansion.
The corresponding wave functions, however, can give substantially different
$\Delta S(k)$ and $\Delta\rho({\bf r})$.

As a general trend, localization of the density redistribution
upon excitation increases progressively from single-CSF to TDDFT 
and BSE toward the CIS result, with BSE giving a result close to CIS in most cases.
There are also rare instances of nuclear configurations where
disorder is sufficient to localize the excitation already at the single-CSF level, 
and others where BSE and TDDFT give relatively extended density redistributions and
localization appears only with the CIS wave function.
The sensitivity to the electronic-structure treatment 
therefore depends strongly on the local molecular environment.

At the highest density, where the exciton becomes larger than the simulation 
cell~\cite{suppmat,Gorelov_2024}, both CIS and BSE become increasingly dominated 
by a single particle-hole transition. In this regime the differences between
the QMC {\em J1CSF} and {\em JCIS} descriptions of $\Delta S(k)$ 
and $\Delta\rho({\bf r})$ become smaller.

Differences among the methods are also found in the transition dipole moments 
and excitation-induced forces, with
BSE following the configuration dependence of the CIS results
more closely than TDDFT.

Overall, our results show that excitation energies alone do not provide a 
sufficient measure of the quality of an excited-state wave function. 
The relatively small differences among QMC excitation energies contrast 
with the substantial changes in exciton localization and the associated 
optical and nuclear response when the CIS expansion is introduced. 
Differences in exciton localization persist even after imaginary-time projection, 
highlighting the importance of the multideterminant trial wave function. 
A correct description of localization is also essential for assessing 
finite-size corrections to excitation energies and obtaining reliable 
predictions in the thermodynamic limit. 
The CIS expansion provides a systematic way to construct such wave functions 
without prior knowledge of the relevant particle-hole transitions. 
Our work thus extends the scope of QMC beyond excitation energies, 
allowing a direct characterization of exciton localization and its 
consequences for optical and nuclear response.

\section*{Acknowledgments}
We thank Vitaly Gorelov and Carlo Pierleoni for discussions and for providing the nuclear configurations
from Ref.~\onlinecite{Gorelov_2024} used here,
Victor Wen-zhe Yu and Giulia Galli for their help in using the WEST code,
and Maurizia Palummo and Daniele Varsano for discussions.
This work was partially funded by the European Union’s Horizon 2020 research and innovation programme 
under grant agreement No.~952165 (TREX Center of Excellence in Exascale Computing).
We acknowledge the use of the Dutch national e-infrastructure with the support of the SURF Cooperative (grant No.~NWO-2025.003), and of the cluster ORFEO at the AREA Science Park in Trieste.

\section*{Data availability}
The codes that generate the data shown in this work are openly available~\cite{CHAMP,PySCF,west,Giannozzi2009}.

\bibliography{cis3}

\end{document}